# Measuring the frequency response of optically pumped metal-clad nanolasers

## CHI XU, WILLIAM E. HAYENGA, MERCEDEH KHAJAVIKHAN[†], AND PATRICK LIKAMWA*

*CREOL, The College of Optics and Photonics, University of Central Florida, Orlando, Florida 32816, USA*
[†]*mercedeh@creol.ucf.edu*
*\*patrick@creol.ucf.edu*

**Abstract:** We report on our initial attempt to characterize the intrinsic frequency response of metal-clad nanolasers. The probed nanolaser is optically biased and modulated, allowing the emitted signal to be detected using a high-speed photodiode at each modulation frequency. Based on this technique, the prospect of high-speed operation of nanolasers is evaluated by measuring the *D*-factor, which is the ratio of the resonance frequency to the square root of its output power ($f_R/P_{out}^{1/2}$). Our measurements show that for nanolasers, this factor is an order of magnitude greater than that of other state-of-the-art directly modulated semiconductor lasers. The theoretical analysis, based on the rate equation model and finite element method simulations of the cavity is in full agreement with the measurement results.

## 1. Introduction

In short-distance communications, such as data swapping, intra-chip, and chip-to-chip interconnects, optical circuits exhibiting higher interconnect density and lower power dissipation have a clear advantage over their electronic counterparts [1-3]. Short-reach optical communications increasingly demand light sources with sufficient output power, small size, low cost, high efficiency, stable single-mode operation, and the capability to be directly modulated at high-speeds [4]. In the past few decades, vertical-cavity surface emitting lasers (VCSELs) have been widely developed and used in 1-100Gb/s communication systems due to their low cost, robust operation, and high performance [5, 6].

In recent years, metal-clad nanolasers have been considered as potential light sources for high-resolution imaging and high-speed communications [7-11]. Their appeal is rooted in some of their intrinsic characteristics; for example, because of the metallic shell, the mode can become fully confined within an ultra-small volume, thus enabling dense integration on chip, while the high coupling efficiency of the spontaneous emission into the lasing mode (large *β*-factor) can lead to low thresholds [12, 13]. In addition, these lasers tend to show a sparse modal spectra and operate in a single-mode fashion due to their miniature size. Altogether, the combination of low power consumption and high modulation speed in metal-clad nanolasers may be used to effectively address some of the main challenges of optical communications where reducing energy per bit is pursued [14-17].

Despite the anticipation regarding the high-speed performance of metallic nanolasers, this aspect has so far remained largely unverified experimentally. In this paper, we bridge this gap by measuring the intrinsic frequency response of metallic nanolasers at low pump power levels [18-20]. To achieve this goal, we build a virtual instrument test set which performs a point-to-point measurement at discrete modulation frequencies. We then use this information to extrapolate the response of nanolasers at higher pump levels, where the speed is expected to surpass our measurement capabilities.

The paper is organized as follows. In section 2, a rate equation model for these type of metallic nanolasers, where the spontaneous emission coupling factor into the lasing mode is large, is introduced. This system of equations is subsequently used to derive the frequency response of the lasers. In section 3, we consider a disk-shape metallic nanolaser and calculate its modal parameters to be used in the rate equations. The simulation finds the eigenfrequencies using a three-dimensional finite element method (FEM) tool. We then discuss in detail the fabrication process in section 4, and describe our custom-made modulation response measurement setup in section 5. In section 6, we characterize and report the intrinsic frequency response of a nanolaser at various pump powers in order to evaluate and predict its high-speed performance. Finally, section 7 concludes the paper.

## 2. Nanolaser rate equations

The rate equations for a nanolaser system are given in Eqs. (1) and (2) [21]. Here, we use the relationship between the stimulated and spontaneous emission rates in order to avoid confusion arising from the treatment of group velocity in such arrangements.

$$\frac{dn_c}{dt} = \frac{\eta_i I}{q} - \frac{F\beta}{\tau_{sp}n_{sp}}n_p n_c - \frac{F}{\tau_{sp}}n_c - \frac{n_c}{\tau_{nr}}, \tag{1}$$

$$\frac{dn_p}{dt} = \frac{F\beta}{\tau_{sp}n_{sp}}n_p n_c + \frac{F\beta}{\tau_{sp}}n_c - \frac{n_p}{\tau_p}, \tag{2}$$

where $n_c$ and $n_p$ are the number of carrier pairs and photons respectively, $\eta_i$ is the current injection efficiency, $I$ is the injection current, q is the elementary charge, $\tau_p$ is the photon lifetime, $\tau_{sp}$ and $\tau_{nr}$ are the multiple quantum wells (MQW) spontaneous and non-radiative recombination lifetimes, respectively. The population inversion factor is defined as $n_{sp} = f_c(1 - f_v)/(f_c - f_v)$, where $f_c$ and $f_v$ are the Fermi-Dirac functions, describing the occupation probabilities in the conduction and valence bands. The Purcell factor, $F$, indicates the cavity-enhanced spontaneous emission rate relative to that of the MQW material, and $\beta$ is the spontaneous emission coupling factor, representing the ratio of the spontaneous emission that is coupled into the lasing mode. These parameters are described in further detail in [7, 22].

The modulation characteristics can be derived by applying a small signal analysis on Eqs. (1) and (2), considering time harmonic variables in the form of ($n_c \rightarrow n_{c_0} + \Delta n_c e^{i\omega t}$, $n_p \rightarrow n_{p_0} + \Delta n_p e^{i\omega t}$, $I \rightarrow I_0 + \Delta I e^{i\omega t}$). The frequency response of a laser system can be written as $H(\omega) = \omega_R^2/(\omega_R^2 - \omega^2 + i\gamma\omega)$, where $\omega_R$ is the relaxation resonance angular frequency and $\gamma$ is the damping factor:

$$\omega_R^2 = \frac{F\beta}{\tau_{sp}n_{sp}\tau_p}n_p + \frac{F\beta}{\tau_{sp}\tau_p}, \tag{3}$$

$$\gamma = \frac{F\beta}{\tau_{sp}n_{sp}}n_p + \frac{F}{\tau_{sp}} + \frac{1}{\tau_{nr}}. \tag{4}$$

The 3-dB cut-off frequency ($f_{3dB}$) is defined as the frequency where the magnitude of the impulse response function reaches half of its DC value, i.e. $|H(\omega)|^2 = 1/2$. When damping is weak, the cut-off frequency, $f_{3dB}$, is approximately proportional to the relaxation resonance frequency, $f_R$, through $f_{3dB} \approx 1.55 f_R$. Clearly, $\omega_R^2$ can be enhanced by increasing $n_p$ or by improving the spontaneous emission rate (larger Purcell factor).

## 3. Cavity simulation and laser performance evaluation

From the above rate equations, some of the cavity related parameters ($\tau_p$, $F$, and $\beta$) play pivotal roles in determining the modulation bandwidth of nanolasers. In order to quantify these parameters for a given nanolaser, we first simulate the electromagnetic properties using an FEM tool. The structure of the metal-clad laser is schematically illustrated in Fig. 1(a). This device is comprised of a cylindrical gain medium consisting of six InGaAsP quantum wells ($d_2 = 200$ nm) protected by a 10 nm InP capping layer and has a radius of $R = 760$ nm. At a temperature of 77 K, the relative permittivities of the gain medium and InP used in the simulation is $\varepsilon_g = 10.56$ and $\varepsilon_{InP} = 9.81$, respectively [9]. The gain is covered by a 50 nm $SiO_2$ layer, $\varepsilon = 2.1$, that serves as a plug and also prevents the formation of plasmonic modes at that interface. A 40 nm air aperture is incorporated below the gain in order to increase the mode confinement while also providing a means to optically pump the laser and to collect its emission. Besides this aperture, the device is entirely covered with silver ($\varepsilon_{Ag} = -135 - 0.4i$ [23]), enabling an ultra-small and localized mode.

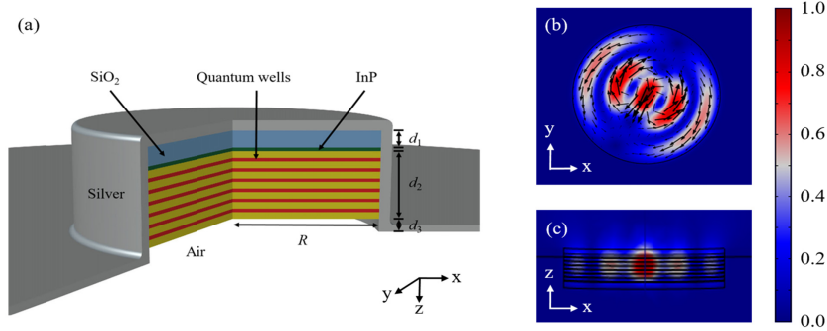

Fig. 1. (a) Schematic of the metal-clad nanolaser; (b) and (c) cross-sections of the cavity mode $TE_{13}$ profile $|E|$.

The above structure is simulated using the Wave Optics module of a commercial FEM package (COMSOL Multiphysics) where the cavity modes are found by an eigenfrequency solver. Among all the possible cavity modes, we are interested only in those overlapping with the gain spectrum of the active material. It is generally expected that the laser oscillations occur predominantly in the cavity modes with the highest net gain. Our simulation shows that for the structure under study, the plasmonic modes (having large radial electric field components towards the metal interface) exhibit smaller quality factors ($Q = f_r/(2f_i)$ for eigenfrequency $f = f_r + if_i$) when compared to the photonic-type modes with electric fields predominately in the azimuthal direction. In particular, for the above nanodisk cavity, the $TE_{13}$-like mode shows the lowest threshold pump level when considering the lineshape of the gain system. Figures 1(b) and 1(c) display the top as well as the cross-sectional views of the normalized electric field of the $TE_{13}$ mode with the eigenfrequency of $f = 2.10 \times 10^{14} + i1.34 \times 10^{10}$. The quality factor, $Q$, of this mode is on the order of ~780, which results in a photon lifetime of $\tau_p = \lambda Q/(2\pi c) \sim 0.6$ ps, where the wavelength in vacuum is $\lambda = 1428$ nm. From simulations, the Purcell factor is estimated to be $F = Q\lambda^3 \xi_{avg}/(4\pi^2 n^3 V_{eff}) = 4.42$, where $\xi_{avg}$ describes locations and orientations of random dipoles in relation to the modal field over the gain region, and $V_{eff}$ is the effective cavity volume. The spontaneous emission coupling factor $\beta$ is found to be 0.193, calculated as the ratio of the Purcell factor of the lasing mode to the summation of the Purcell factors for all modes within the gain spectrum. The spontaneous emission lifetime for the quantum wells is $\tau_{sp} = 1/(BN) = 14.4$ ns, determined by the rate equation model at steady state, where the radiative recombination coefficient B is taken as $8.9 \times 10^{-10}$ cm$^3$/s and $N$ is the carrier density [8, 24].

The frequency response and the modulation bandwidth of the above metallic laser can be estimated by inserting the related parameters into Eqs. (3) and (4). Figure 2(a) depicts the modulation bandwidth as a function of the injection current for a metal-clad nanolaser. This device exhibits an ultra-low threshold of $I_{th} = 2\pi c n_{sp} q[1 + \tau_{sp}/(F\tau_{nr})]/(\lambda \beta Q) \sim 66$ μA.

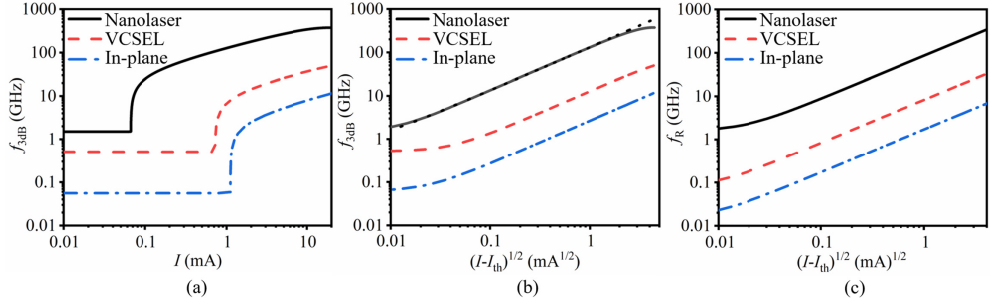

Fig. 2. Simulated modulation bandwidths of a metallic nanolaser, a VCSEL, and an in-plane laser. (a) 3-dB bandwidth vs. injected current; (b) 3-dB bandwidth vs. $\sqrt{I - I_{th}}$; (c) relaxation resonance frequency vs. $\sqrt{I - I_{th}}$.

The modulation bandwidth $f_{3dB}$ is expected to linearly increase with the square root of the pump current until the strong damping saturates it. The maximum modulation bandwidth is found to be 374.8 GHz at $I = 19.5$ mA. However, effects such as gain saturation and heating can limit the attainable modulation frequency, long before reaching this value [25]. Nevertheless, a modulation bandwidth near 100 GHz is well feasible under a sub-mA injection current. This figure also compares the nanolaser response to that of VCSELs [5] and in-plane lasers [26]. The parameters used in simulations of all three devices are provided in Table 1. The modulation bandwidth of the VCSEL is taken to be ~30 GHz at $I = 10$ mA, while most in-plane lasers show much lower speeds. The modulation current efficiency factor (MCEF), a figure of merit to compare the high-speed performance of semiconductor lasers is defined as $f_{3dB}/\sqrt{I - I_{th}}$. This factor is used to assess the prospect of various lasers in terms of their high-speed performance. Figure 2(b) shows $f_{3dB}$ versus $\sqrt{I - I_{th}}$ for the nanolaser, VCSEL, and in-plane laser. From this plot, an MCEF of 132 GHz/mA$^{1/2}$ can be found for the nanolaser, which is more than ten times larger than that of VCSELs (~12.7 GHz/mA$^{1/2}$). Finally, Fig. 2(c) displays how the relaxation resonance frequencies evolve as a function of $\sqrt{I - I_{th}}$. The slope of these curves, determines the $D^*$-factor ($D^* = f_R/\sqrt{I - I_{th}}$) [27], which is found to be 86.4 GHz/mA$^{1/2}$ for the nanolaser, 8.3 GHz/mA$^{1/2}$ for the VCSEL, and 1.7 GHz/mA$^{1/2}$ for the in-plane laser.

**Table 1. Parameters used in the simulated modulation performances.**

| Parameters | Nanolaser | VCSEL | In-plane laser |
|---|---|---|---|
| $\lambda$ (nm) | 1428 | 1550 | 980 |
| $F$ | 4.42 | 1 | 1 |
| $\beta$ | 0.193 | $2.25 \times 10^{-4}$ | $8.69 \times 10^{-5}$ |
| $Q$ | 780 | 4230 | 5325 |
| $\tau_{nr}$ (ns) | 0.07 | 18 | 20 |
| $\tau_{sp}$ (ns) | 14.4 | 0.28 | 3.32 |
| $n_{sp}$ | 1.003 | 1.5 | 1.13 |
| $\eta_i$ | 80% | 80% | 80% |

## 4. Fabrication

The fabrication steps involved in implementing the metal-clad nanolaser are depicted in Fig. 3. The gain medium consists of 200 nm thick $In_{x=0.734}Ga_{1-x}As_{y=0.57}P_{1-y}$ (20 nm)/ $In_{x=0.56}Ga_{1-x}As_{y=0.938}P_{1-y}$ (10 nm) MQW grown on an InP substrate. On a meticulously cleaned piece of a wafer, a layer of negative electron beam resist hydrogen silsesquioxane (HSQ) is spun. After 10 minutes of prebaking at 180°C, an EBPG 5000+ e-beam lithography system is used to write the pattern, which is then developed in tetramethylammonium hydroxide (TMAH). After exposure, the HSQ is converted to an SiO$_2$-like material that serves as a mask to transfer the pattern to the MQW structure by reactive ion etching (RIE) with gas proportions of $H_2$:$CH_4$:Ar = 40:4:20 SCCM. Next, a layer of silver with a thickness of 2 μm is deposited on the sample using e-beam evaporation. For mechanical handling, we flip the wafer and bond it to a glass slide. Lastly, the sample is immersed in HCl to remove the InP substrate.

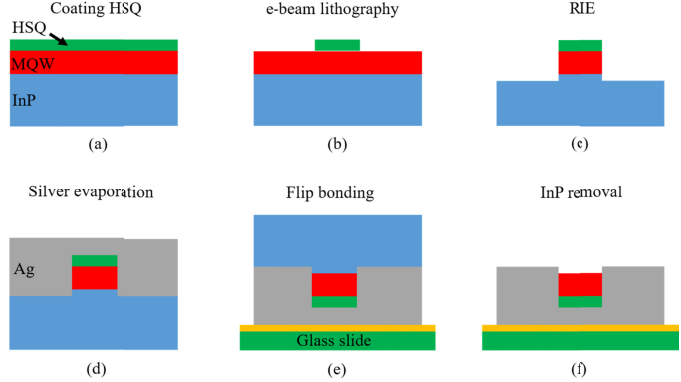

Fig. 3. Main fabrication steps for implementing the nanolaser. (a) HSQ is spun onto the wafer; (b) the pattern is defined by means of e-beam lithography; (c) the disk structure is formed by reactive ion etching; (d) the cavities are coated with silver via e-beam deposition; (e) the wafer is flipped and bonded to a glass slide; (f) lastly the InP is removed by HCl immersion.

## 5. Measurement station

The schematic of the test setup is shown in Fig. 4. A 1064 nm Nd:YAG laser operating in CW mode is used to pump the metal-clad nanolaser. In addition, light generated by a 1310 nm semiconductor laser is passed through an electro-optic modulator with a 15 GHz modulation bandwidth and is used as a modulation signal which is consequently amplified and polarization corrected using a semiconductor optical amplifier (SOA) and a polarization controller (PLC). A dichroic beam splitter is used to combine the 1310 nm signal with the 1064 nm pump beam. A 50X long-working-distance objective is used to focus the pump onto and to collect the emission by the nanolasers. A polarization beam splitter (PBS) reflects the combination of pump and signal beams onto the sample and directs the output emission to the designated measurement tools. A long-pass optical filter is used to block the 1064 nm pump and 1310 nm signal from the detector. For obtaining a higher output power, the metal-clad nanolaser is inserted into a cryostat and is cooled down to 77 K using liquid nitrogen ($LN_2$). The laser can work at room temperature, albeit with a lower output power. However, in this measurement, we used a cooled device in order to have better signal to noise ratio. A confocal microscope scheme is used to align the nanolaser at the center of the pump beam. The laser emission with the modulated output signal is captured by a high frequency photodiode (Newport 818-BB-35) connected to an RF spectrum analyzer (HP 8560). A HP 8720 vector network analyzer (VNA) generates the RF signal which is amplified using a Mini-Circuits wideband amplifier ZVA-213+ before being applied to the electro-optic modulator. In order to measure the frequency response of the metallic nanolaser at low output powers, a virtual instrument test set is constructed. During the measurement, the response is determined point-by-point for discrete frequencies instead of directly measuring the system scattering parameter $S_{21}$ using the vector network analyzer. At each modulation frequency point, the VNA generates an RF signal, while the spectrum analyzer runs a single sweep with a 5 kHz range centered at this same frequency. The analyzer searches for the peak amplitude of the detected RF signal, which is then recorded by the control computer. The response of the nanolaser is obtained by repeating the above procedure over a range of frequencies, and normalizing the result to the system response in the absence of the device under study. Due to the small size of the cavity, the modulated signal is relatively weak. To solve this problem, we use the above method to lower the noise floor of the RF spectrum analyzer. However, the electromagnetic interference increases at higher frequencies, limiting the capability of the measurement technique to ~ 9 GHz.

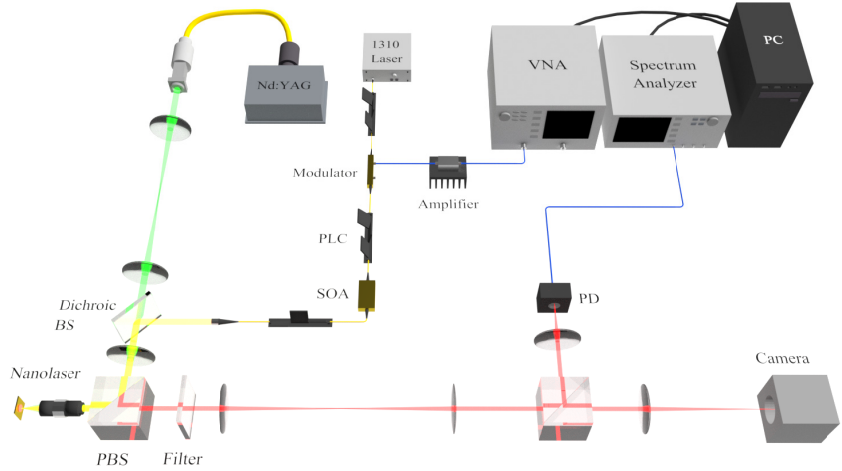

Fig. 4. Schematic of the frequency response test setup. The Nd:YAG laser (1064 nm) is used as the pump, while the modulated signal is provided by a laser operating at 1310 nm.

## 6. Experimental results

The probed metal-clad nanolaser has a cavity radius of $R = 760$ nm and lases at a wavelength of 1428 nm. Figure 5(a) shows the laser output intensity profile captured by an infrared camera. Polarization resolved emission is further provided in Figs. 5(c) and 5(e) and is compared to that of the simulated emission profiles of a quasi-linear $TE_{13}$-type mode (Figs. 5(d) and 5(f)). The measured light-light curve of the metal-clad nanolaser is displayed in Fig. 5(g), clearly showing a threshold pump intensity of 8 µW/µm².

The nanolaser is operated within the linear region of the light-light curve, where the frequency response is measured in a point-by-point fashion at various output powers (Fig. 6(a)), as described in Section 5. At relatively low pump levels, the measured response exhibits the expected bandwidth broadening associated with an increasing output power ($P_{out}$). The characteristics of the intrinsic response $f_R$, $f_{3dB}$, and $\gamma$ are determined by fitting the measured curves with respect to the theoretical expression $|H|^2 = f_R^4/[f_R^4 + f^4 - 2f_R^2 f^2 + (\gamma f/2\pi)^2]$, shown as solid curves in Fig. 6(a).

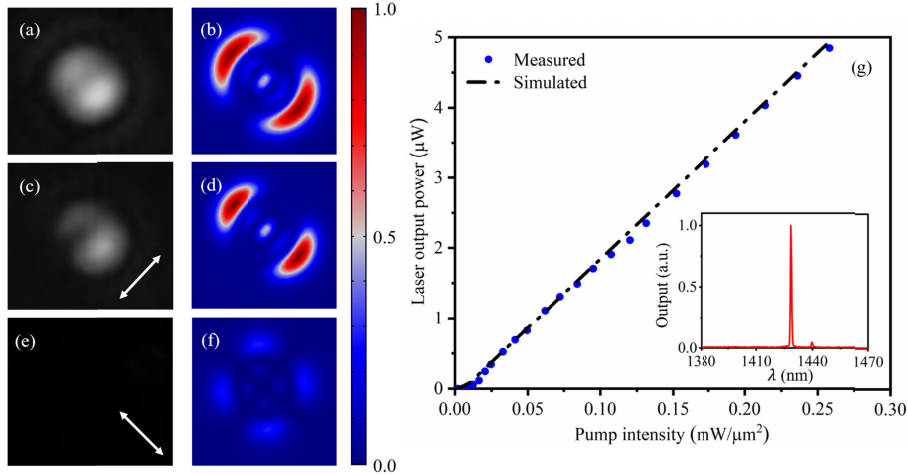

Fig. 5. Laser characteristics. (a) Laser intensity profile; (b) Simulated normalized $|E|^2$ of the laser output; Measured (c), (e) and simulated (d), (f) orthogonal polarization components of the laser emission (the polzariation direction is indicated by the arrow). To accurately capture the emission intensity, the PBS reflecting the pump is exchanged with a non-polarizing beam splitter; (g) The light-light curve of the probed metallic nanolaser. Experimental results are shown as blue dots and the simulated estimation is depicted as black dashes and dots. The inset shows the lasing spectrum at a pump power level of 130 µW.

For evaluating the high-speed performance of the nanolaser, $f_R^2$ vs. $P_{out}$ is plotted in Fig. 6(b). In addition, a simulated $f_R^2$ vs. $P_{out}$ (dotted line) from Fig. 2(c) is provided. Here a dependence of $P_{out} = (I - I_{th})\eta_i \eta_o h\nu/q$ is used, where the optical output efficiency $\eta_o$ is close to 1 for this short cavity. The slope of linear fit of the experimental results is 11.6 GHz$^2$/μW. A figure of merit to characterize the intrinsic modulation response is the $D$-factor, defined as $D = f_R/\sqrt{P_{out}}$ [27]. The $D$-factor for the above nanolaser is found to be ~107.5 GHz/mW$^{1/2}$. To the best of our knowledge, this is the highest value reported for semiconductor lasers to date. Considering an electrical pumping scheme for nanolasers as reported in [25], and a current injection efficiency of $\eta_i = 80\%$, the MCEF estimated from the experimental results is 130 GHz/mA$^{1/2}$, which is more than an order of magnitude greater than current high speed VCSELs [5].

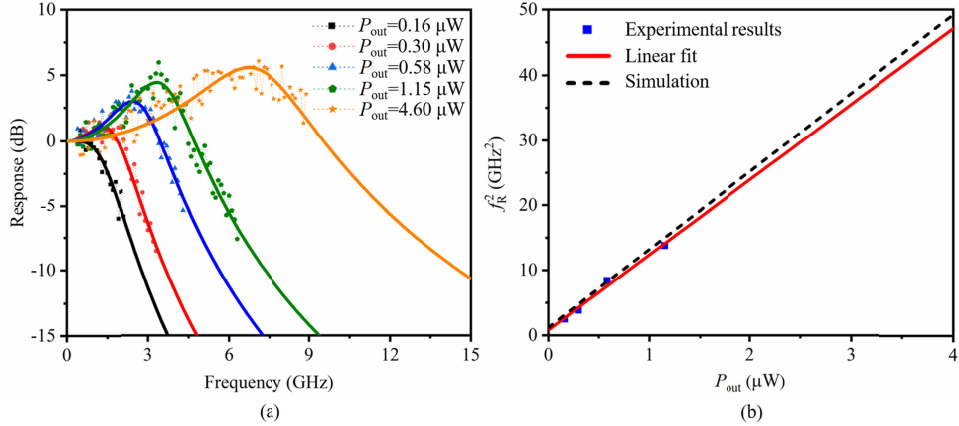

Fig. 6. (a) Measured frequency responses of a metallic nanolaser at various output powers; (b) Relaxation resonance frequency squared vs. output power.

## 7. Conclusion

The intrinsic frequency response of metal-clad nanolasers has been theoretically evaluated using a rate equation model and experimentally investigated employing a point-by-point characterization method. Our numerical simulation results show that the modulation bandwidth rapidly increases with an MCEF of 132 GHz/mA$^{1/2}$. In the experiment, we measure the modulation response at relatively low pump levels (9 GHz at $P_{in} = 0.25$ mW/μm$^2$), where we find a $D$-factor of 107.5 GHz/mW$^{1/2}$. This value is an order of magnitude larger than that reported for other semiconductor lasers. The theoretical simulations of the metallic nanolaser also suggest a maximum modulation frequency greater than 300 GHz – something that yet to be experimentally verified. These exceptional features are in large attributed to the cavity enhanced spontaneous emission rate and the high $\beta$-factor, which enable high-speed operation, even at low bias levels. We believe our results could help pave the way for designing the next generation of fast and efficient lasers.

### Funding